\documentclass[11pt]{article}

\usepackage[final]{acl}

\usepackage{times}
\usepackage{latexsym}

\usepackage[T1]{fontenc}

\usepackage[utf8]{inputenc}

\usepackage{microtype}

\usepackage{inconsolata}

\usepackage{graphicx}

\usepackage{hyperref}       
\usepackage{url}            
\usepackage{booktabs}       
\usepackage{amsfonts}       
\usepackage{tcolorbox}      
\usepackage{nicefrac}       
\usepackage{microtype}      
\usepackage{xcolor}         
\usepackage{tikz} 
\usepackage{wrapfig}  
\usepackage{pifont}
\usepackage{float}

\usepackage{tabularx} 
\usepackage{ragged2e} 
\usepackage{multirow}

\usepackage[shortlabels]{enumitem}

\title{Position: LLM Watermarking Should Align  Stakeholders’ Incentives for Practical Adoption}

\author{
  Yepeng Liu$^{\dagger}$, ~~ Xuandong Zhao$^{\ddagger}$, ~~ Dawn Song$^{\ddagger}$, ~~ Gregory W. Wornell$^{\S}$, ~~ Yuheng Bu$^{\dagger}$ \\
  $^{\dagger}$UC Santa Barbara ~~~~ $^{\ddagger}$UC Berkeley ~~~~ $^{\S}$Massachusetts Institute of Technology \\
  \texttt{\normalsize \{yepengliu, buyuheng\}@ucsb.edu, \{xuandongzhao, dawnsong\}@berkeley.edu,  gww@mit.edu}
}

\begin{document}
\maketitle

\begin{abstract}
Despite progress in watermarking algorithms for large language models (LLMs), real-world deployment remains limited. 
We argue that this gap stems from misaligned incentives among LLM providers, platforms, and end users, which manifest as three key barriers: competitive risk, detection-tool governance, and attribution issues.
We revisit three classes of watermarking through this lens. \emph{Model watermarking} naturally aligns with LLM provider interests, yet faces new challenges in open-source ecosystems. \emph{LLM text watermarking} offers modest provider benefit when framed solely as an anti-misuse tool, but can gain traction in narrowly scoped settings such as dataset de-contamination or user-controlled provenance. \emph{In-context watermarking} (ICW) is tailored for trusted parties, such as conference organizers or educators, who embed hidden watermarking instructions into documents. If a dishonest reviewer or student submits this text to an LLM, the output carries a detectable watermark indicating misuse. This setup aligns incentives: users experience no quality loss, trusted parties gain a detection tool, and LLM providers remain neutral by simply following watermark instructions. We advocate for a broader exploration of incentive-aligned methods, with ICW as an example, in domains where trusted parties need reliable tools to detect misuse. More broadly, we distill design principles for incentive-aligned, domain-specific watermarking and outline future research directions. Our position is that the practical adoption of LLM watermarking requires aligning stakeholder incentives in targeted application domains and fostering active community engagement.

\end{abstract}

\section{Introduction}

The widespread adoption of large language models (LLMs) \cite{grattafiori2024llama, yang2024qwen2} has intensified concerns about misuse, as these models increasingly produce human-like outputs. To enhance attribution and accountability, watermarking has emerged as a key approach.
This includes LLM text watermarks, which embed imperceptible signals in generated content to identify AI output, and model watermarks, which encode signatures directly into model parameters to trace unauthorized use. Together, these techniques aim to support content provenance, protect intellectual property (IP), and promote trust in AI \cite{zhao2024sokwatermark, liu2024survey, pan2024markllm}.

Common \emph{LLM text watermarking} strategies include embedding watermarks by perturbing the next-token prediction distribution \cite{kirchenbauer2023watermark, zhao2023provable, liu2024adaptive, liu2024semantic}, and employing pseudo-random sampling \cite{gumbel2023, christ2023undetectable, kuditipudi2023robust, hu2023unbiased, he2024universally}.
These methods have demonstrated effectiveness in terms of detectability, robustness, and text quality.  Beyond text, popular \emph{model watermarking} techniques include watermarking during fine-tuning~\cite{xu2023instructions, xu2025mark, zhaosurvey, nasery2025scalable, xu2024instructional}, resisting model extraction via APIs~\cite{zhao2023protecting, sander2024watermarking, panaitescu2025can, zhao2022distillation}, and protecting IP datasets~\cite{jovanovic2024ward, liu2025dataset, wei2024proving}. \emph{However, despite substantial research efforts and many proposed techniques, real-world adoption of watermarking remains limited}.

In this paper, we explore the key reasons hindering the broader adoption of LLM watermarking, including \emph{competitive risks}, governance challenges of \emph{detection tools}, and \emph{attribution} issues. Among these, we identify the \textbf{lack of aligned incentives} as the most fundamental barrier. Without clear benefits for LLM providers, users, and other stakeholders, even well-designed watermarking algorithms backed by regulation may struggle to achieve broad real-world adoption.

\textbf{This position paper advocates that, to advance the adoption of LLM watermarking, we should look beyond technical performance and consider the broader ecosystem of stakeholders that influence design and deployment. We believe effective watermarking systems need to be tailored to specific application domains, grounded in clear threat models, and aligned with the interests of model providers, platforms, and users.} Some existing concerns can be resolved when the use case is well defined and stakeholder incentives are properly addressed.
\textbf{Our position does not oppose existing or future regulatory efforts, but emphasizes that regulation alone is insufficient to drive widespread adoption.} Effective uptake of watermarking requires well-aligned incentives, not solely by mandates. Moreover, since well-designed regulations and enforcement typically progress slowly, an incentive-aligned approach can be deployed immediately without requiring international consensus or regulatory oversight, helping to bridge the gap until effective global regulations are established.

To support our position, the remainder of the paper is organized around three types of watermarking techniques applied in different use cases: \emph{model watermarking}, \emph{LLM text watermarking}, and the newly proposed \emph{in-context watermarking}.
For each, we analyze the specific incentive model and assess whether stakeholder incentives are aligned to support real-world adoption. Where alignment exists, we draw analogies from traditional watermarking systems to highlight relevant lessons and design principles. Where alignment is lacking, we explore how existing techniques can be adapted or repurposed for alternative application domains to better align with stakeholder interests.
Specifically, 
\begin{itemize}
[leftmargin=*,itemsep=0em]
    \item \emph{Model watermarking} enables LLM developers to trace unauthorized uses without affecting end-user experience. By directly protecting the provider’s IP at no cost to users, it aligns naturally with provider incentives and encounters minimal adoption friction, though it may face new challenges in open-source ecosystems.

    \item \emph{LLM text watermarking} provides modest direct incentive for LLM providers when used solely to mitigate misuse (e.g., academic dishonesty) and can even push users toward unwatermarked LLMs. However, its value rises when repurposed to serve provider interests, such as filtering self-generated data to prevent model collapse, or helping users safeguard confidential material.
    
    
    \item \emph{In-context watermarking (ICW)} embeds a watermark via instructions in the input prompt. It aligns provider incentives by placing control of watermarking in the hands of trusted parties, such as conference organizers or educators.
    Existing preliminary exploration demonstrates the effectiveness of ICWs on the most advanced LLMs, highlighting the promise of this direction for broad real-world deployment.
\end{itemize}

\section{Issues of Current LLM Text Watermarking System}
\label{sec:issues}
\vspace{-0.5em}
Despite steady progress in LLM watermarking, real-world adoption remains limited. To date, only Google’s SynthID has been deployed on Gemini Web and mobile endpoints~\cite{gloaguen2024black}.
As Scott Aaronson noted at the ICLR 2025 Workshop on GenAI Watermarking, his proposal to integrate Gumbel-max watermarking into OpenAI’s models was ultimately not adopted \cite{gumbel2023}.
We next outline three key barriers hindering real-world adoption.

\textbf{\ding{172} Competitive Risks.}
Implementing LLM text watermarking exposes early adopters to immediate competitive risks. If a single company adopts watermarking, users who fear being labeled as AI-generated, or who dislike the change, can switch to other LLM providers without watermarking or use open-source models locally. This dynamic places responsible companies at a disadvantage when prioritizing AI safety. In short, with the current usage of LLM text watermarking, the market discourages, rather than rewards, its adoption.

\begin{figure*}[t!]
    \centering
    \includegraphics[width=1\linewidth]{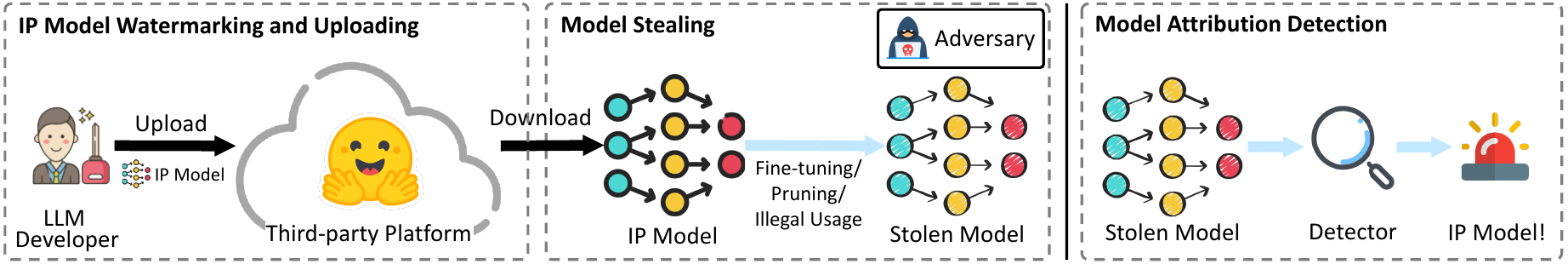}
    \vspace{-2em}
    \caption{Example of model watermarking: an adversary fine-tunes, prunes, or illegally uses a protected model, and the LLM developers detect the unauthorized model.}
    \label{fig:overview_model_watermarking}
    \vspace{-1.2em}
\end{figure*}

\textbf{\ding{173} Governance of Detection Tools.}
Current LLM text watermarking schemes typically rely on a secret key during detection, but distributing and managing this key in practice presents significant challenges. In most cases, LLM providers are expected to pair a watermarked LLM API with a detector API\footnote{Google recently released a public API for detecting SynthID watermarking in AI-generated images. See: \url{https://blog.google/technology/ai/google-synthid-ai-content-detector/}
}. However, if the detection API or the key is made fully public, it opens the door to adversarial probing and spoofing, introducing additional security risks~\citep{pang2024no,jovanovic2024watermark}. On the other hand, restricting access, such as limiting it to educational institutions or select third parties, requires costly centralized infrastructure and raises concerns about fairness, transparency, and antitrust issues.

\textbf{\ding{174} Lack of Attribution Issues.}
Many high-quality LLM outputs reflect substantial human effort, including careful prompt engineering and post-editing, not just AI’s contribution. Labeling such text as ``AI-generated'' (or worse, ``AI-authored'') using current watermarking tools overlooks the user’s input and unfairly penalizes those, especially non-native speakers, who rely on LLMs for language assistance and refinement~\cite{liang2023gpt,cooper2023report}. A multi-bit watermark that enables fine-grained attribution, rather than a simple binary indicator, may offer an effective solution.

\section{Model Watermarking}
\vspace{-0.5em}

We begin by examining model watermarking setups (Figure \ref{fig:overview_model_watermarking}) and their incentive model, showing how stakeholder interests may align to support wider adoption. Drawing on lessons from traditional digital watermarking systems such as iTunes, we highlight key design principles while acknowledging the challenges unique to open-source platforms.

\textbf{Related Work.}
Our focus is on model watermarking to protect model weights against misuse, such as theft, pruning, and unauthorized fine-tuning. Watermarking against other threats, like model stealing via distillation, typically requires distinct strategies and is discussed in Appendix~\ref{app:related}. In open-source LLMs, developers distribute weights via platforms like Hugging Face or GitHub. Developers or platforms can embed watermarks in model weights for later verification.  One method is model backdooring~\cite{xu2023instructions, xu2025mark, zhaosurvey, lou2023trojtext, xue2023trojllm}, where the LLM developer fine-tunes the model to associate a secret trigger (e.g., a specific token sequence) with a predefined output (e.g., a fixed phrase) that would not occur naturally. In this way, the model behaves as intended under normal usage but `reveals' the watermark (abnormal behavior) with the trigger presented. Another method is model fingerprinting \cite{nasery2025scalable, xu2024instructional, russinovich2024hey, zhang2024vtune, cai2024utf, yamabe2024mergeprint}, where the model is fine-tuned on key-response pairs. Unlike backdoors, the responses are not a single secret phrase or overtly abnormal behavior, but subtle preference patterns across many queries that enable model identification.

\subsection{Incentive Model}

\begin{figure}[h]
    \centering
    \vspace{-1em}        \includegraphics[width=0.7\linewidth]{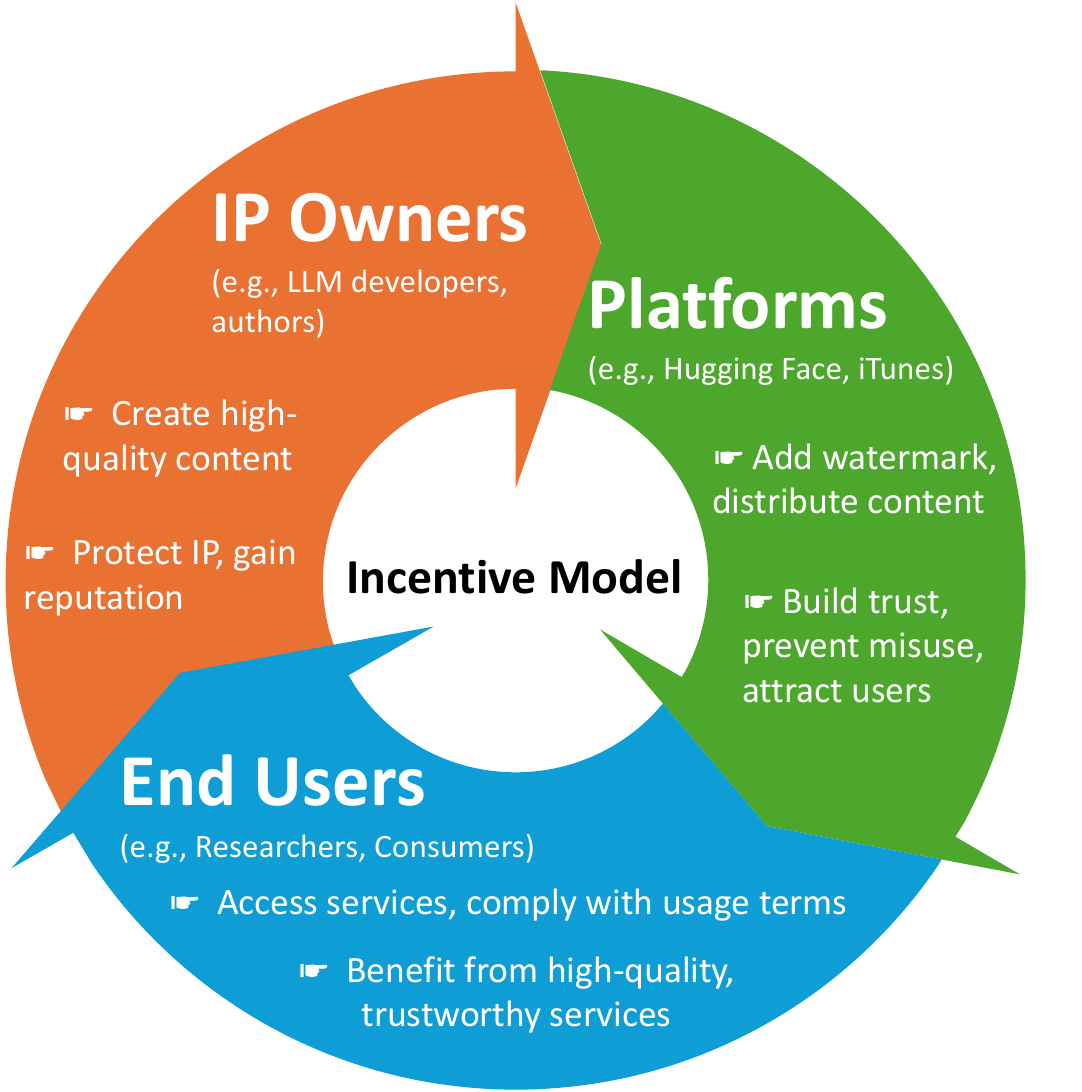}
    \vspace{-0.5em}
    \caption{Incentive model for model watermarking among IP Owners, platforms, and users.}
\label{fig:model_watermarking_threat_model}
\vspace{-0.5em}
\end{figure}

The incentive model, illustrated in Figure~\ref{fig:model_watermarking_threat_model}, involves three entities: IP owners, platforms, and end users, whose interests can align through model watermarking. In the open-source scenario, developers upload pretrained models to platforms such as Hugging Face to gain visibility and community adoption. Yet permissive licenses leave them exposed to adversaries who might rebrand, resell, or redistribute the weights without credit. In this context, model watermarking acts as a lightweight attribution mechanism, allowing developers to trace usage, assert ownership, and deter misuse without disrupting legitimate adoption by normal users.

\begin{table*}[h!]\scriptsize
\centering
\setlength{\tabcolsep}{9pt}
\caption{Analogy between model watermark and traditional digital watermark (e.g., iTunes). } 
\label{tab:model_ditial_comparison}
\vspace{-1em}
\begin{tabular}{ccccccc}
\toprule
 & \textbf{IP Owner} & \textbf{Platform} & \textbf{Adversary} & \textbf{Goal} & \textbf{Technique} & \textbf{Detector Owner}   \\ \midrule
\textbf{Model Weights}   & LLM Developer & Hugging Face  & Model Thief     &   IP Protection   & Watermark/Fingerprint     &  Platform or IP Owner                   \\
\textbf{Digital Goods} &  Content Creator & iTunes & Media Pirate  & IP Protection     & Watermark/Metadata   & Platform              \\ \bottomrule
\end{tabular}
\vspace{-2em}
\end{table*}

In the Model-as-a-Service (MaaS) setting (e.g., ChatGPT), LLM developers host their own APIs without an intermediary platform. Growing usage boosts their visibility, reputation, and subscription revenue. Adversaries, however, can erode this value by extracting large volumes of data, distilling the models, or deploying unauthorized replicas. And LLM developers have the incentive to adopt model watermarking (detailed in Appendix~\ref{app:related}) to trace such misuse and safeguard their IP.

From this perspective, model watermarking closely parallels traditional digital watermarking systems used in domains such as music and visual art, where watermarks may be applied either by platform (e.g., Hugging Face or iTunes) or by IP owners to protect IP and deter unauthorized use. A more detailed comparison is provided in Table~\ref{tab:model_ditial_comparison}.

Because model watermarking directly safeguards an IP owner’s core asset, it offers incentive for adoption and may see broad uptake. Potential deployment scenarios include tracing unauthorized model distillations and flagging research misconduct, as illustrated by recent plagiarism incidents~\cite{xinhua2024stanford}, where proprietary data was used as watermark to support claims of unauthorized model use.

\subsection{What iTunes Taught Us}

Given the similarity in incentive structures, the evolution of iTunes offers valuable insight into how model watermarking could develop over time.
The iTunes Store, launched in 2003, illustrates a shift from restrictive enforcement toward user-centric, forensic watermarking.
During the FairPlay era (2003-2007), tracks were sold as encrypted .m4p files playable only on Apple devices, with embedded metadata identifying the purchaser. While DRM (Digital Rights Management) restricted playback, the watermark itself was visible with any MP4 parser.
In 2007, Apple removed DRM from newly purchased songs, but retained the user-specific forensic tag in .m4a file. Users gained the freedom to copy and play music anywhere, while a persistent identifier continued to link leaked files to the original purchaser. By 2009, the entire iTunes catalog was DRM-free, yet every download still carried a visible watermark, ensuring accountability without compromising user experience.

iTunes succeeded not because of sophisticated watermarking technology but because the watermark was embedded in a service users already preferred. Songs were cheap, easy to buy with one click, and synced seamlessly to iPods, putting user experience first. Over time, the watermark shifted from being restrictive (blocking unauthorized playback) to forensic (providing traceability if content leaked). It was stored as visible metadata, so no special tools were needed to detect it.

The case offers a useful lesson for model watermarking on platforms like Hugging Face: watermarking added by the platform should not degrade user experience, be transparently verifiable, and be integrated into a product offering that users actively prefer over unwatermarked alternatives.

\subsection{New Challenges}

While threat models share similarities, platforms like Hugging Face introduce new challenges distinct from traditional watermarking contexts. Unlike the music industry's standardized MP4 format, the model ecosystem lacks a unified file structure. Frequent model transformations (fine-tuning, repackaging, quantization, pruning) and open-source model distribution complicate watermarking \cite{dai2025seal, xu2025copyright, guo2025invariant, zhang2024emmark, fernandez2024functional, li2024double, li2023plmmark}. Consequently, watermarks demand high robustness to persist through format conversions and weight modifications. Moreover, watermark needs to extend beyond model weights to associated assets like datasets, embeddings, and outputs, crucial for model training and deployment. (Dataset watermarking methods are discussed in Appendix~\ref{app:related}).

In open-source ecosystems, the potential for malicious platforms incentivizes developers to embed watermarks themselves, contrasting with the iTunes model, where the platform, not creators, typically applies watermarks. Additionally, while model watermarking aims to protect IP owners, it is susceptible to misuse. A dishonest model owner could fabricate infringement claims by withholding or falsifying watermark keys, falsely asserting that their watermark appears in another model. To mitigate such abuse, platforms like Hugging Face could serve as neutral arbiters. This role involves collecting and storing verified watermark metadata and maintaining clear ownership records. These governance needs highlight a crucial future direction: designing model watermarking systems for technical robustness, accountability, and trust within collaborative ecosystems. Moreover, to prevent platforms from becoming ``trust bottlenecks'', their neutrality could be secured in two ways. Economically, a platform’s value as a two-sided market depends on its reputation; biased verification would drive developers to competitors, creating a self-correcting market incentive for fairness. Technically, platforms should adopt auditable, open-standard protocols to ensure the community can independently verify outcomes and prevent black-box governance.

\vspace{-0.2em}

\section{LLM Text Watermarking}
\vspace{-0.2em}
We now shift focus to watermarking the text generated by LLMs. In particular, we explore how the incentives of LLM providers may not align with the broader goals of AI safety, preventing the widespread adoption of text watermarking.

\textbf{Related Work.}
LLM text watermarking typically embeds a watermark by manipulating the decoding process of LLMs, including logits perturbation and pseudo-random sampling. The detailed discussion can be found in Appendix \ref{app:related}.

\subsection{Misaligned Incentive}

\begin{figure}
    \centering
    \includegraphics[width=0.7\linewidth]{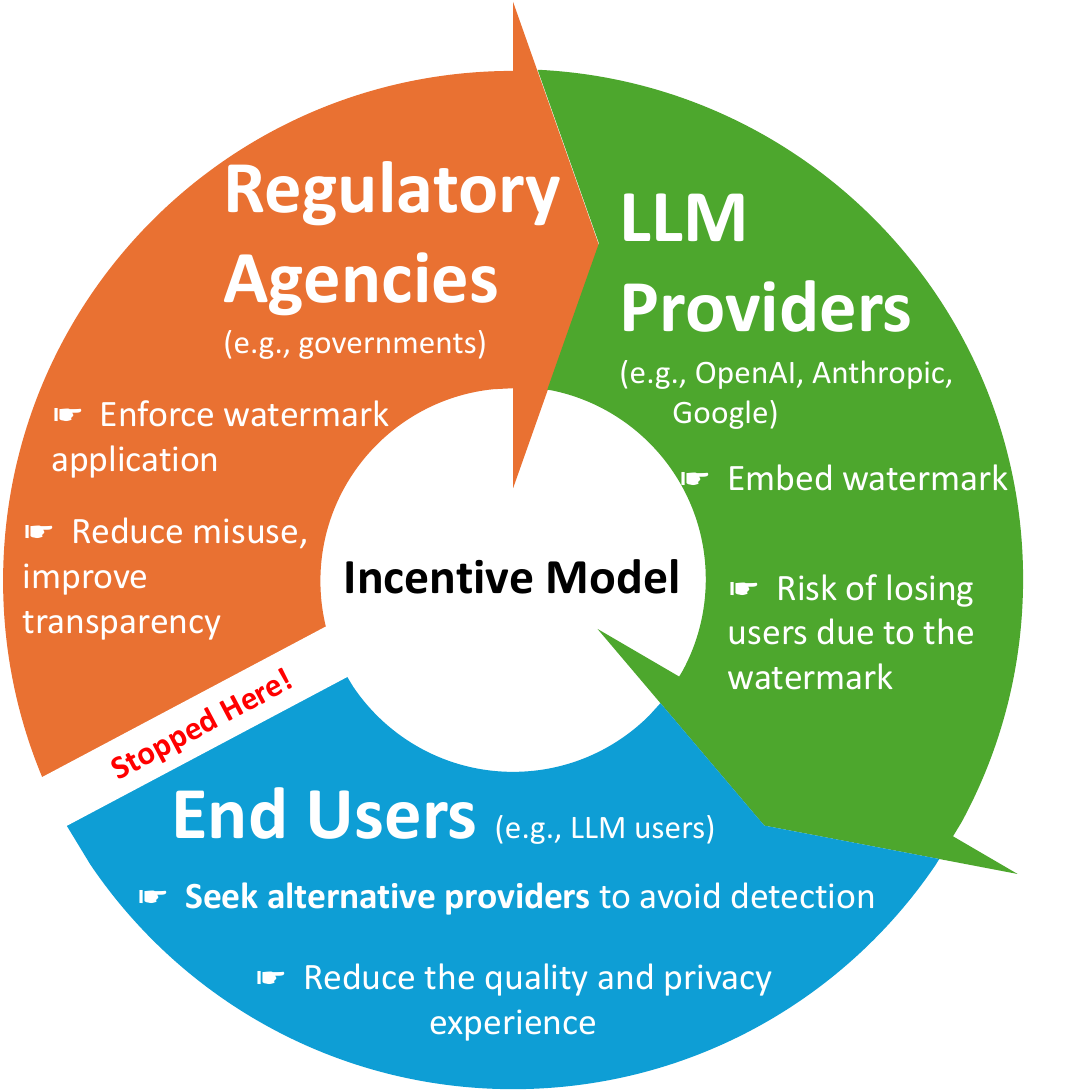}
    \vspace{-0.5em}
    \caption{Broken Incentive Model for LLM Text Watermarking: Users may switch to unwatermarked models, undermining both the LLM provider’s interests and the intended goal of reducing misuse. }
    \label{fig:llm_text_watermarking_threat_model}
    \vspace{-1em}
\end{figure}

We analyze the incentive model (Figure~\ref{fig:llm_text_watermarking_threat_model}) for the LLM text watermarking when used to prevent LLM misuse. In this setting, LLM providers embed watermarks in all generated text to prioritize AI safety.
While this approach can help identify some adversarial uses, it also introduces a significant trade-off: users who object to having their generated content labeled or traceable may simply switch to unwatermarked models offered by other providers. As a result, while watermarking may help mitigate certain types of misuse and serve the public interest, it provides little direct benefit to LLM providers. This underscores a fundamental misalignment between the incentives of LLM developers and the goals of broader AI safety efforts.

One may expect that regulatory agencies may play a critical role in curbing AI misuse and internalizing its externalities. However, such efforts often fall short due to \emph{jurisdictional limitations}, \emph{competition from unregulated regions}, and the \emph{widespread availability of locally deployed open-source models}.

Moreover, current LLM text watermarking techniques are limited to establishing the provenance of AI content, and they do not directly detect misuse. 
Hence, it should not be viewed as a universal fix for AI abuse. Instead, watermarking should
be deployed in targeted settings where the incentives of stakeholders naturally align. Below, we present two use cases where existing techniques offer clear benefits to the stakeholders, showing how a shift in application domain can mitigate several of the challenges noted in Section \ref{sec:issues}.

\subsection{Use Case: Watermarking Benefits LLM Developers}

\begin{figure*}[t]
    \centering
    \includegraphics[width=1\linewidth]{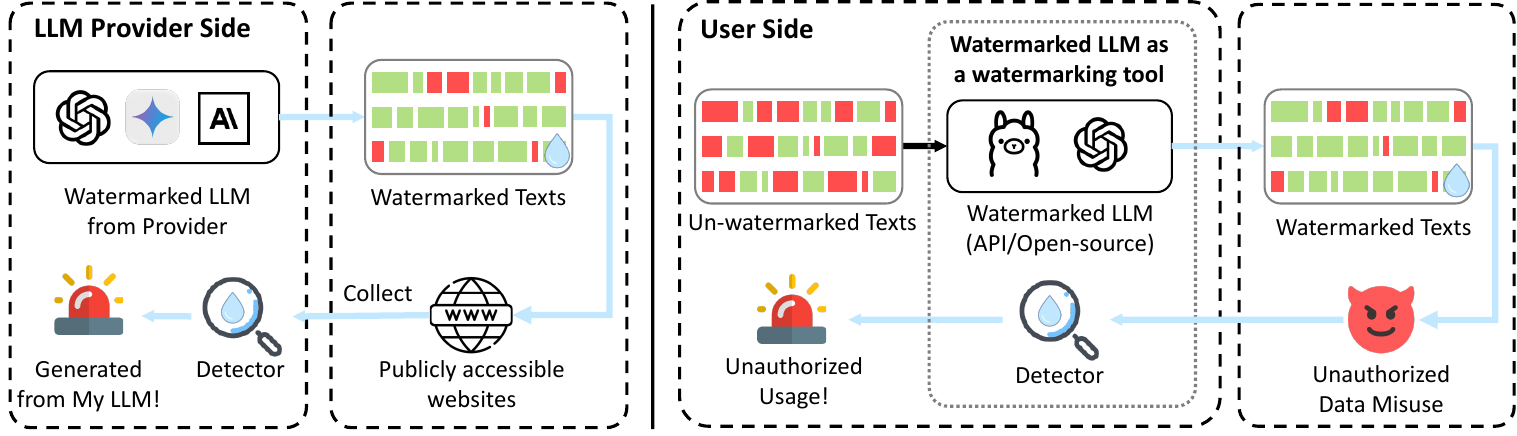}
    \vspace{-1.8em}
    \caption{Illustration of two LLM text watermarking use cases. Left: Watermarking implemented by LLM Provider to detect self-generated data; Right: Watermarking implemented by the users to safeguard the user's document. } 
    \vspace{-1.3em}
    \label{fig:overview_llm_watermarking}
\end{figure*}

LLM text watermarking can be used by LLM developers to filter out texts generated by their models when collecting training data, as shown in Figure~\ref{fig:overview_llm_watermarking} Left. This helps prevent model collapse caused by training on synthetic data, as highlighted in~\cite{shumailov2024ai}. 
In this use case, watermarking serves as a data-curation tool, improving corpus quality rather than detecting misuse, which directly benefits developers by enhancing the performance of future LLMs.

In this case, all three issues discussed in Section \ref{sec:issues} can be effectively resolved. 
\ding{172} Users are unlikely to be aware of the watermark, and it does not impact their experience, e.g., using the undetectable watermark~\cite{christ2023undetectable}, eliminating competitive risk.
\ding{173} Since this watermark is used exclusively by developers, there is no need to make the detection tool public.
\ding{174} The watermark can be extended to encode multi-bit information, such as the model version and timestamp, to support fine-grained attribution.

Some text watermarks subtly shift the LLM token distribution. Because text watermark persists after unauthorized distillation~\cite{gu2023learnability,sander2024watermarking}, they provide a covert, model-level signature for ownership verification and theft tracing. Therefore, LLM text watermarking can also serve as model watermarking, deterring proprietary model extraction as discussed in Appendix~\ref {app:related}.

We note that these applications of LLM text watermarking benefit developers but rest on the assumption that providers act honestly. In settings like Chatbot Arena~\cite{chiang2024chatbot}, where users evaluate model outputs blindly, a dishonest provider could exploit watermarking to identify their own model's responses, effectively bypassing the blind evaluation and unfairly boosting their leaderboard position. ~\cite{min2025improving,singh2025leaderboard}. Future research and incentive design are needed to ensure watermarking serves as a tool for data integrity, not a means of deception.

\subsection{Use Case: Watermarking Benefits LLM Users}
LLM text watermarking can also protect end users' confidential material, as shown in Figure~\ref{fig:overview_llm_watermarking} Right. 
In one scenario, the users run an open-source LLM locally to paraphrase passages from a sensitive report; before returning the rewritten text, the model embeds an invisible watermark tied to the source document, much like a hidden ``CONFIDENTIAL'' stamp in a PDF. In a second scenario, a university partners with a commercial LLM provider to deliver summaries of restricted documents via a controlled-access API. Each response is likewise stamped with a covert mark derived from a secret key held only by the research team.
In both situations, the watermark works as a digital signature that lets owners trace any unauthorized sharing.

In both scenarios, user-side watermarking addresses three of the main obstacles that limit the broader adoption of LLM text watermarking. 
\ding{172} Because the watermark is user-requested, it can be offered as an optional LLM feature that attracts users rather than deterring them.
\ding{173} The secret key and the lightweight detector remain under user control, removing the need for a public API and mitigating misuse or antitrust concerns.
\ding{174} Multi-bit metadata (e.g., user ID, timestamp) enables reliable attribution by authenticating the watermark and identifying its source.
The incentive structure is therefore well aligned: users gain a lightweight confidentiality enforcement tool, while LLMs (open-source or API-based) preserve output quality.

\section{In-context Watermarking}

Most existing LLM text watermarking methods focus on determining whether a piece of text was generated by an AI model, rather than addressing the misuse of LLMs in specific, high-stakes contexts. However, many real-world scenarios, such as a conference organizer trying to detect AI-written peer reviews or a teacher seeking to identify LLM-generated homework, involve content created outside these trusted parties' control. In these cases, existing LLM text watermarking approaches, which rely on modifying the model generation process, are difficult to apply. This highlights the need for alternative strategies that can operate in user-driven workflows.

\begin{figure}[h]
    \centering
    \includegraphics[width=\linewidth]{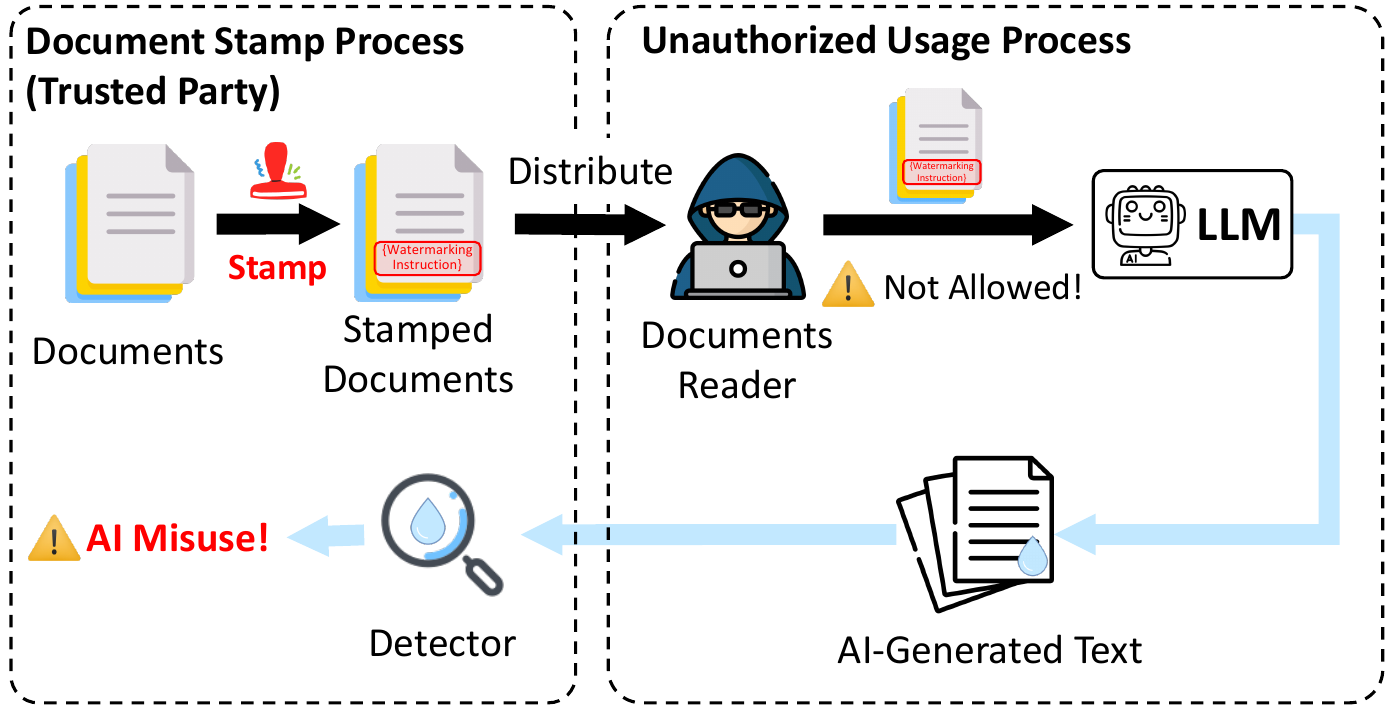}
    \vspace{-1.8em}
    \caption{Overview of In-Context Watermark.}
    \label{fig:overview_in_context_watermark}
    \vspace{-0.5em}
\end{figure}

One promising approach is to modify the LLM input. Since many lazy reviewers (or students) paste documents directly into LLMs for summarization or drafting, documents can be embedded with imperceptible ICW instructions. These signals subtly influence the LLM’s output, allowing downstream detection without altering the model or disrupting the reviewer’s workflow.

This strategy motivates a new form of LLM text watermark, \emph{ICWs}, and its application in sensitive settings such as peer reviews or student homework. ICWs leverage LLM's in-context learning~\cite{dong2022incontextsurvey, brown2020fewshot} and instruction-following abilities~\cite{zhou2023instruction, mu2023caninstruction} to embed detectable signals into generated text. By inserting carefully designed watermarking instructions into the prompt, LLMs can produce watermarked outputs, enabling reliable detection without modifying the model itself. The effectiveness of ICWs, including detection performance, robustness, and text quality, has been demonstrated empirically in existing works \cite{liu2025incontext, zhong2024watermarking, rao2025detecting}. Results indicate that, with well-designed watermarking instructions, ICW achieves strong performance across both proprietary and open-source models. Detailed experimental results are reported in these studies.

\textbf{Related Work.} The existing research on ICW is limited. Specifically, \citet{liu2025incontext} investigates four ICW strategies: adding invisible characters, altering lexical choices, modifying initials, and using acrostics. The study finds that ICW effectiveness improves with LLM capabilities and shows strong performance in detection accuracy, robustness, and text quality across both in-process generation and indirect prompt injection scenarios (e.g., paper reviews).
\citet{zhong2024watermarking} proposes a method that uses a prompting LLM to generate context-aware watermarking instructions and a marking LLM to embed these watermarks into the generated text. A classifier is then trained to detect the presence of the watermark. \citet{rao2025detecting} designs a method specifically for detecting LLM-generated reviews. The approach injects a prompt into manuscripts that guides LLMs to include predefined patterns in the generated reviews, such as random start phrases, technical terms, or fake citations.

\subsection{Exploration of Simple ICWs}

The threat model (Figure \ref{fig:overview_in_context_watermark}) of ICW applied to peer review setting involves three entities: authors, reviewers, and conference organizers. Authors submit papers, and reviewers evaluate them. Conference organizers aim to maintain the integrity of the review process by identifying dishonest reviewers who violate policy by uploading submissions to LLMs for automated review. 
Organizers can covertly embed a watermarking instruction into the manuscript, for example, by using `white text' (text colored the same as the background) within the PDF. If a reviewer inputs such a manuscript (containing the hidden instruction) into an LLM, the generated review may carry a detectable watermark.
While authors could embed their own prompts to identify AI-generated reviews, this poses a conflict of interest, they may falsely accuse negative reviews of being AI-generated. Therefore, watermarking should be administered by conference organizers, who act as trusted parties.

A similar threat model applies to student homework, where the instructor embeds the watermarking instruction. Unlike authors in peer review, teachers do not have a conflict of interest, making the approach simpler to implement and manage in educational settings.

\textbf{A simple example.} As an illustrative example of ICW, we present Initials ICW, as introduced in \cite{liu2025incontext}. The Initials ICW scheme embeds a watermark by encouraging LLMs to bias the initial letters of words in generated text to a subset (green letters) of English alphabet letters. An abbreviated watermarking instruction is shown below.

\begin{tcolorbox}[
  top=2pt,          
  bottom=2pt        
]
\#\# Watermarking Instruction:

Maximize the use of words starting with letters from \{green\_letter\_list\}.
\end{tcolorbox}

Initials ICW increases the proportion of green initial letters in generated text, and detection is by computing the z-statistic over the frequency of green initial letters in the suspect text. It demonstrate effectiveness, especially for advanced LLMs with strong instruction-following capabilities.

In Appendix \ref{app:icw_experiments}, we present brief experimental results on advanced LLMs to illustrate the performance of ICWs and demonstrate their potential for practical deployment.

\subsection{Incentive Model}

\begin{figure}[h]
\vspace{-1em}
    \centering
    \includegraphics[width=0.7\linewidth]{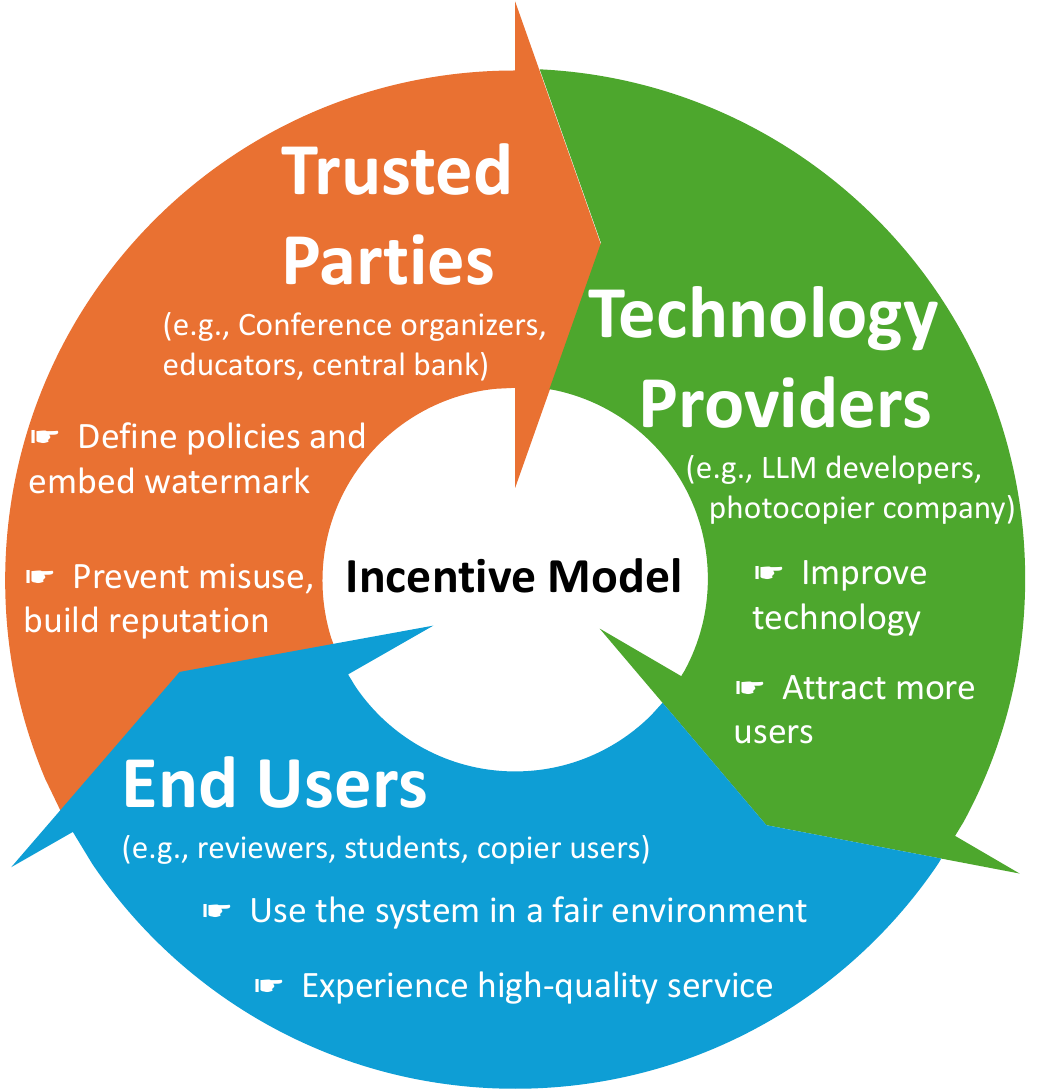}
    \vspace{-0.5em}
    \caption{Incentive model for model watermarking among trusted parties, technology providers, and users.}
    \label{fig:icw_watermarking_threat_model}
    \vspace{-1em}
\end{figure}

\begin{table*}[t]\scriptsize
\centering
\setlength{\tabcolsep}{6pt}
\caption{Analogy between ICW and EURion constellation.} 
\vspace{-1em}
\begin{tabular}{ccccccc}
\toprule
 & \textbf{Trusted
Parties} & \textbf{Technology
Providers} & \textbf{Adversary} & \textbf{Goal}  & \textbf{Response Mechanism}  \\ \midrule
\textbf{ICW}   &Organizer/Teacher& LLM Developer  &  Dishonest LLM User   & Trace AI Misuse         &Embed ICW            \\
\textbf{EURion} &Central Bank  & Photocopier Company  &  Counterfeiter  & Prevent Money Counterfeit    & Stop Service               \\ \bottomrule
\end{tabular}
\label{tab:analogy_icw_eurion}
\vspace{-1.3em}
\end{table*}

The incentive model (Figure~\ref{fig:icw_watermarking_threat_model}) of ICWs differs sharply from that of model and LLM text watermarking. For ICWs, the watermark is inserted not by the LLM provider or the end user to protect their interests but by a trusted third party, such as conference organizers and educators, whose goal is to identify dishonest LLM use. The LLM provider’s only requirement is to support reliable instruction following so that embedded watermarking instructions are executed reliably. This aligns incentives across stakeholders: organizers obtain higher-quality, human-authored reviews, instructors uphold academic integrity, normal users are unaffected; only dishonest behavior is flagged.

ICW aligns the incentives of different parties, thereby avoiding the usual deployment barriers. \ding{172} Because trusted parties embed the watermark in the prompt, LLM providers face no competitive risk and need not fear user loss. \ding{173} Governance of the detection tool is straightforward: trusted third parties alone hold the keys and detectors, eliminating conflicts of interest. 
\ding{174} attribution is unambiguous: a detected mark directly links the LLM-generated text to a specific reviewer or student, enabling reliable enforcement of policy.

\subsection{Rethinking the Analogy}

The incentive model of ICW mirrors that of the EURion constellation used in modern banknotes. Central banks embed a subtle, machine-readable pattern in the currency; printers and photocopiers recognize the pattern to prevent money counterfeiting, while everyday users remain unaffected. Likewise, ICW lets a trusted party embed an imperceptible signal that LLMs dutifully follow, enabling reliable post-hoc detection of misuse without degrading the normal user experience.

Following this analogy (Table \ref{tab:analogy_icw_eurion}), it is conceivable that LLM providers could collaborate with conference organizers or universities to design specific patterns, similar to the EURion constellation, that elicit predefined model behaviors. For example, when a confidential document containing such a pattern is provided as input, the model could recognize its sensitivity and either embed an imperceptible watermark in the output or refuse to generate a response, while avoiding internalizing the content during training. This leads to a proactive misuse prevention method in LLM. Specifically, unlike the EURion constellation, embedding an imperceptible watermark in the output is less noticeable to users, whereas halting generation may signal the protection mechanism and invite workarounds or even denial-of-service attacks.

As an emerging paradigm, ICWs face several open challenges that warrant further investigation. These include their reliance on LLM instruction-following capabilities, robustness to sophisticated attacks, and ethical and transparency considerations. Finally, for clarity, we summarize the differences among model watermarking, LLM text watermarking, and in-context watermarking with respect to primary deployment actors, use cases, incentive alignment, and failure modes in Table \ref{tab:watermarking-comparison}.

\section{Discussion of Future Direction}
\label{app:future_direction}

\textbf{Toward Principled Multi-Bit Watermarking.} Several of our use cases require richer provenance for reliable attribution, e.g., document ID, user ID, or timestamps, driving the need for multi-bit watermarking. Existing methods typically bolt on simple coding schemes in an ad hoc way~\cite{he2025distributional, yoo2023robust, qu2024provably, boroujeny2024multi, wang2023towards}. A more principled approach is to view watermarking as an information-embedding problem~\cite{chen2001quantization,martinian2005authentication} and apply information-theoretic tools to establish fundamental limits. Recent efforts~\cite{he2025distributional} have begun exploring this direction, which can inform the design of optimal coding strategies for more robust and efficient multi-bit watermarking.

\textbf{Benchmarking ICW as a Measure of Instruction-Following Capability.} An appealing aspect of ICW is that its effectiveness improves with more capable LLMs. As better instruction-following directly enhances watermark performance, it creates a natural incentive for LLM developers to support third-party watermarking use cases. To advance this direction, future work should establish standardized benchmarks to evaluate a model’s ability to embed ICWs, positioning this as a new metric for instruction following. Building such datasets and evaluation protocols is only a starting point, but it will help guide both research and industry toward more reliable, user-controlled watermarking solutions.

\clearpage

\section*{Limitations}

There are several alternative perspectives to the positions we present in the paper. Some researchers and policymakers advocate regulatory mandates to ensure consistent deployment and accountability. Because the harms of AI misuse are widely distributed and hard to monetize, market incentives alone are insufficient. As a result, top-down regulation is considered the most reliable path to achieving broad and timely adoption of watermarking technologies. Moreover, the emergence of anti-detection markets may challenge incentive alignment and hinder efforts to detect AI misuse, as LLM providers could also have incentives to weaken or bypass watermarking. Our framework mainly analyzes primary alignment goals and simplifies the diverse and often contested incentive structures in the real-world. In future research, we will explore more deeply how these stakeholders navigate complex strategic trade-offs across diverse regulatory environments. There is no single method that is a panacea. For those alternative perspectives and concerns, we have a more detailed discussion in Appendix \ref{app:alternative_view}.

\bibliography{ref}

\clearpage

\appendix

\section{Other Related Works}
\label{app:related}

\begin{table*}[t]
\centering
\scriptsize
\caption{Comparative summary of watermarking settings across deployment actors, goals, incentive alignment, and potential failure modes.}
\label{tab:watermarking-comparison}
\begin{tabularx}{\textwidth}{l >{\RaggedRight\arraybackslash}X >{\RaggedRight\arraybackslash}X >{\RaggedRight\arraybackslash}X >{\RaggedRight\arraybackslash}X}
\toprule
\textbf{Settings} & \textbf{Primary Deployment Actors} & \textbf{Primary Goal/Use Case} & \textbf{Incentive Alignment} & \textbf{Failure Modes} \\ \midrule
Model Watermarking & LLM developers or Platforms (e.g., Hugging Face) & IP protection & \textbf{Aligned:} Protect the provider's core asset without degrading user experience. & Transparency, disputes over ownership proofs \\ \addlinespace

LLM Text Watermarking & LLM provider (e.g., OpenAI, Google) & Provenance of AI-misuse & \textbf{Misaligned:} Create competitive risk for model providers; users may switch to unwatermarked models. & Market rejection, key management \\ \addlinespace

In-Context Watermarking & Trusted third parties (e.g., Conference organizers) & Provenance of AI-misuse & \textbf{Aligned:} Trusted parties gain detection tools; model providers remain neutral; users get fair services. & Dependence on model's instruction-following ability \\ \bottomrule
\end{tabularx}
\end{table*}

\begin{figure*}[t]
    \centering
    \includegraphics[width=1\linewidth]{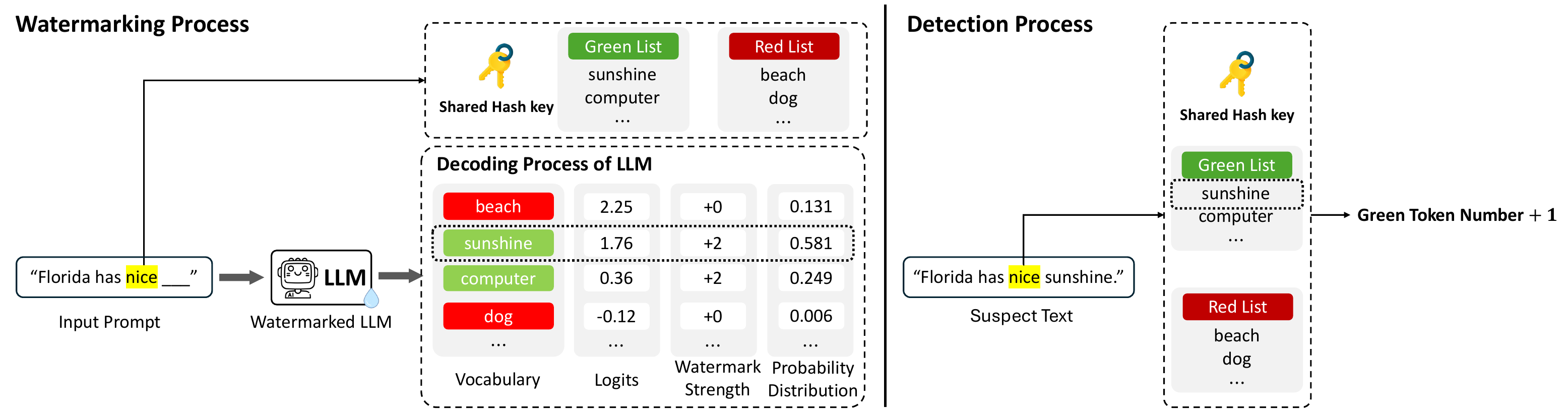}
    \caption{Illustration of Green/Red list LLM text Watermarking~\cite{kirchenbauer2023watermark}. }
    \label{fig:example_llm_watermarking}
    \vspace{-1em}
\end{figure*}

\textbf{LLM Text Watermarking.} LLM text watermarking typically embeds a watermark by manipulating the decoding process of LLMs \cite{he2025distributional, li2024robust, li2025statistical, liu2023unforgeable, zhang2025cohemark, zhu2024duwak, fu2024gumbelsoft, xu2024rlwatermark, huo2024token, hou2023semstamp, ren2023robust, dathathri2024scalable, giboulot2024watermax, fernandez2023three, yoo2023robust, qu2024provably, ghosal2023towards, chakraborty2023possibilities, ji2025overview, liu2024image}, including logits perturbation \cite{liu2024adaptive,liu2024semantic, huang2025watermarking, lee2023wrote} and pseudo-random sampling \cite{gumbel2023, he2024universally, chen2025improved, zhao2025permute}. Specifically, Figure~\ref{fig:example_llm_watermarking} illustrates the green/red list watermarking method~\cite{kirchenbauer2023watermark}, which partitions the LLM vocabulary into green and red token sets. The model is then subtly biased toward generating green tokens by modifying the output logits during sampling.
\cite{gumbel2023} uses the Gumbel-Max trick to pseudo-randomly sample the next token during the text generation. Moreover, in addition to manipulating the decoding process, \cite{xu2024rlwatermark} trains a paired LLM and detector to embed and detect watermarks collaboratively. \cite{bahri2024watermark} introduces a black-box approach that, at each generation step, generates several candidate $n$-grams and selects the one with the highest hash-based score. Unlike in-process watermarking, which embeds the watermark during generation, post-hoc watermarking modifies text after it has been generated~\cite{brassil1995electronic, por2012unispach, sato2023embarrassingly, rizzo2016content, yang2023blackbox, yang2022tracing, meral2009natural, topkara2006words, an2025defending, an2026reinforcement, chang2024postmark, zhang2024remark, qiang2023natural}. Moreover, some works investigate the quality or alignment problem caused by existing watermarking techniques
\cite{ajith2024downstream, molenda2024waterjudge, verma2025watermarking}.

\textbf{Proprietary LLM Extraction.} For proprietary LLMs, the adversary typically has access only to the model’s output (e.g., text) via its API. These outputs can then be used to label a substitute dataset, which enables the adversary to train a surrogate model. The most common strategy is LLM text watermarking, which involves manipulating the decoding process of proprietary LLMs \cite{kirchenbauer2023watermark, zhao2023provable, gu2023learnability} without requiring additional training. The core idea is that the watermark signal embedded in the model-generated text can be learned by the surrogate model trained on those watermarked outputs, resulting in the surrogate’s outputs also carrying detectable watermark information. Specifically, \cite{zhao2023protecting} injects a secret sinusoidal modulation into the token-generation logits, creating invisible `spectral' signatures in the sequence of chosen tokens. \cite{sander2024watermarking} demonstrates the radioactivity of existing LLM text watermarks \cite{kirchenbauer2023watermark}, showing that when watermarked text is used as fine-tuning data, the watermark signal is transferred to the fine-tuned model.

\textbf{Unauthorized Dataset Misuse.} In addition to model IP infringement, protecting dataset IP is also crucial. Unauthorized users may incorporate proprietary datasets into their model training data or use them in Retrieval-Augmented Generation (RAG) systems \cite{karpukhin2020dense, xiong2020approximate, lewis2020retrieval} without permission \cite{panaitescu2025can, liu2023watermarking, cui2025robust, liu2025mask, li2024seeing, anderson2024my}. The dataset owner usually embeds a backdoor \cite{chaudhari2024phantom, cheng2024trojanrag, chen2024agentpoison} or watermark \cite{liu2023watermarking,jovanovic2024ward, liu2025dataset, wei2024proving} into the dataset for reliable detection. Specifically, \cite{zou2024poisonedrag} introduces a technique that involves inserting crafted malicious content into the dataset, causing retrieval-augmented LLMs to produce a specific, incorrect response to a targeted query. \cite{liu2025dataset} inserts carefully crafted watermarked canaries into the proprietary dataset to detect unauthorized use of the dataset in RAG systems.

\section{Experimental results of ICWs}
\label{app:icw_experiments}
In this section, we present brief experimental results on the detection and robustness performance of ICWs, given that the concept is relatively new. The performance is evaluated under the indirect prompt injection (IPI) setting, where we consider a scenario in which academic conference organizers embed watermarking instructions into submitted manuscripts and then detect if a review is generated by inputting the manuscript to an LLM.

The experiments use ICLR papers from 2020 to 2023 as a dataset. The ICWs are evaluated on \texttt{gpt-4o-mini} and \texttt{gpt-o3-mini}. The performance is evaluated using ROC-AUC, true positive rate at 1\% false positive rate (TPR@1\%FPR), and true positive rate at 10\% false positive rate (TPR@10\%FPR). The robustness is evaluated by paraphrasing the watermarked text using LLMs.

As shown in Table \ref{tab:icw-performance}, ICW performance improves with increasing LLM capability, for example, from GPT-4o-mini to GPT-o3-mini. For advanced models such as GPT-o3-mini, ICWs achieve strong detection performance. Moreover, ICWs, such as Initials, Lexical, and Acrostics ICWs, remain certain robustness even when the watermarked text is completely paraphrased, demonstrating their potential for practical deployment. More extensive experimental results can be found in \citet{liu2025incontext}.

\begin{table*}[t]
\centering
\scriptsize
\setlength{\tabcolsep}{7pt}
\caption{Empirical performance of ICWs: Comparing detection effectiveness and robustness against paraphrasing on indirect prompt injection (IPI) setting.}
\label{tab:icw-performance}
\begin{tabularx}{\textwidth}{l l ccc ccc}
\toprule
\multirow{2}{*}{\textbf{Language Models}} & \multirow{2}{*}{\textbf{Methods}} & \multicolumn{3}{c}{\textbf{Detection (IPI Setting)}} & \multicolumn{3}{c}{\textbf{Robustness (Paraphrase)}} \\
\cmidrule(lr){3-5} \cmidrule(lr){6-8}
 & & \textbf{ROC-AUC} & \textbf{TPR@1\%FPR} & \textbf{TPR@10\%FPR} & \textbf{ROC-AUC} & \textbf{TPR@1\%FPR} & \textbf{TPR@10\%FPR} \\
\midrule
\multirow{4}{*}{\textbf{GPT-4o-mini}} & Unicode ICW & 0.857 & 0.714 & 0.735 & 0.500 & 0.010 & 0.100 \\
 & Initials ICW & 0.620 & 0.006 & 0.076 & 0.616 & 0.000 & 0.070 \\
 & Lexical ICW & 0.889 & 0.054 & 0.564 & 0.887 & 0.048 & 0.556 \\
 & Acrostics ICW & 0.592 & 0.002 & 0.448 & 0.591 & 0.000 & 0.378 \\
\midrule
\multirow{4}{*}{\textbf{GPT-o3-mini}} & Unicode ICW & 1.000 & 1.000 & 1.000 & 0.500 & 0.010 & 0.100 \\
 & Initials ICW & 0.997 & 0.910 & 0.998 & 0.893 & 0.106 & 0.628 \\
 & Lexical ICW & 0.997 & 0.974 & 0.989 & 0.940 & 0.558 & 0.872 \\
 & Acrostics ICW & 0.997 & 0.982 & 0.998 & 0.874 & 0.448 & 0.724 \\
\bottomrule
\end{tabularx}
\end{table*}

\section{Alternative Views and Discussion}
\label{app:alternative_view}

\textbf{Watermarking as A Mandatory Safety Baseline.} Some people argue that waiting until all parties voluntarily adopt watermarks sets the bar too high. Instead, they treat provenance watermarking as a basic feature in the era of genAI, like seat belts or food allergy labels, that should be mandated through policy, not left to voluntary adoption. From this perspective, watermarking is not merely a market feature driven by aligned incentives but a necessary safeguard to protect society from the risks of AI-generated content.

Regulatory momentum supports this view:
\begin{itemize}
[leftmargin=*,topsep=-0.2em,itemsep=-0.2em]
    \item The EU AI Act~\cite{EU_AI_Act_2024} explicitly requires providers to embed machine-readable watermarks in any system that generates or manipulates content, with enforcement set to take effect in 2025.
    \item China’s Cyberspace Administration~\cite{China_GenAI_Measures_2023} has gone further, mandating both visible and invisible watermarks for generative content and requiring platforms to detect and flag unmarked media.
    \item In the U.S., NIST’s 2024 report on synthetic content~\cite{Chandra2024SyntheticContent} frames watermarking as a foundational content authentication tool, recommended even in the absence of strong commercial incentives. 
\end{itemize}

Supporters of this approach argue that because the harms of misuse, such as deepfake-driven misinformation, copyright infringement, are broadly distributed and hard to monetize~\cite{ma2022specialization}, no individual stakeholder has a strong financial reason to act alone. They contend that without regulatory pressure, the market will reward providers who skip provenance controls in favor of speed, cost, or user satisfaction. As a result, they advocate for watermarking mandates backed by penalties and procurement rules, arguing that these top-down mechanisms are more likely to ensure timely and universal adoption than waiting for stakeholder incentives to naturally align.

\noindent\textbf{The Existence of Anti-detection Markets.} The emergence of anti-detection markets may challenge incentive alignment and hinder efforts to detect AI misuse, as LLM providers could also have incentives to weaken or bypass watermarking.

\noindent\textbf{Discussion.} \emph{In this paper, we advocate for the consideration of incentive alignment when designing the LLM watermarking algorithm for broader deployment in the real world.} However, there is no single method that is a panacea. A well-designed regulation may also play an important role in the ecosystem. However, a well-designed regulation usually requires substantial effort, moves slowly, varies across jurisdictions, and is hard to keep pace with the rapid deployment of LLMs. In the meantime, we believe it is important to consider mechanisms such as ICW that can be deployed immediately without requiring international consensus or regulatory enforcement. Moreover, a simple, one-size-fits-all mandate risks overlooking the diverse incentives of different stakeholders. From a policymaking perspective, we believe it is important to account for these diverse incentives when designing watermarking requirements. In particular, effective regulation should aim to align incentives so that watermarking offers clear value not only to providers but also to trusted third parties and end users. By grounding regulation in incentive alignment, policymakers can reduce resistance and increase the likelihood of successful adoption.

Moreover, the existence of anti-detection markets may challenge incentive alignment, as LLM providers could have motives to weaken watermarking. However, this problem also arises under regulatory approaches. Moreover, incentive-aligned methods could better mitigate such risks: universal watermark mandates create a single target for evasion, while incentive-aligned methods apply mainly in high-stakes contexts with limited incentive to bypass detection. Incentive-aligned methods like ICW also align with providers’ goals to improve instruction-following, requiring no universal cooperation, and opposing it would conflict with their broader interests. Because enterprises and educators all benefit from trustworthy watermarking, deliberately undermining it would bring reputational risks, making compatibility with incentive-aligned methods the rational choice.

\end{document}